\documentclass{article}

\usepackage{arxiv}

\usepackage[utf8]{inputenc} 
\usepackage[T1]{fontenc}    
\usepackage{hyperref}       
\usepackage{url}            
\usepackage{booktabs}       
\usepackage{amsfonts}       
\usepackage{nicefrac}       
\usepackage{microtype}      
\usepackage{lipsum}
\usepackage{graphicx}
\usepackage{subfigure}
\usepackage{multirow}
\usepackage{caption}
\title{Explainable Rumor Detection using Inter and Intra-feature Attention Networks}

\author{
  Mingxuan Chen\thanks{This paper has been accepted for publication in TrueFact 2020.} \\
  Department of Electrical and Computer Engineering\\
  Stevens Institute of Technology\\
  Hoboken, NJ 07030 \\
  \texttt{mchen20@stevens.edu} \\
   \And
 Ning Wang \\
  Department of Electrical and Computer Engineering\\
  Stevens Institute of Technology\\
  Hoboken, NJ 07030 \\
  \texttt{nwang7@stevens.edu}  \\
     \And
 K.P. Subbalakshmi \\
  Department of Electrical and Computer Engineering\\
  Stevens Institute of Technology\\
  Hoboken, NJ 07030 \\
  \texttt{ksubbala@stevens.edu}  \\
}

\begin{document}
\maketitle

\begin{abstract}
\label{sec:abs}

With social media becoming 
ubiquitous, information consumption
from this media has also increased. 
However, one of the serious problems
that has emerged with this increase, is the propagation of rumors. 
Therefore, rumor identification is a very critical task with significant implications
to economy, democracy as well as public health and safety. 
We tackle the problem of automated 
detection of rumors in social media 
in this paper by designing a modular 
explainable
architecture that uses both latent 
and handcrafted
features and can be expanded to as many new classes of features as desired. 
This approach will allow the end user to not only determine whether the piece of information on the social media is real of a rumor, but also give explanations on why the algorithm arrived at its conclusion.
Using attention mechanisms, we are able to interpret the relative importance
of each of these features as well as the relative importance of the feature
classes themselves. 
The advantage of this approach is
that the architecture is expandable 
to
more handcrafted features as they 
become available and also to conduct
extensive testing to determine the relative influences of theses features in the final decision.
Extensive experimentation on popular 
datasets and 
benchmarking against eleven 
contemporary algorithms, show that 
our approach
performs significantly better in 
terms of F-score and accuracy 
while also being interpretable.
\end{abstract}


\section{Introduction}

It has been well established that the Internet, especially social 
networks, provides a platform for ``viral" spread of information at 
rates faster than even a fully connected traditional networks 
\cite{doerr2012rumors}. 
Depending on the actors involved, this could either be used for 
societal good or ill. For e.g, a rumor
about an explosion in the White House caused the Dow Jones Industrial Average to immediately plunge and the S\&P 500 was reported to have lost \$136.5 billion in market cap, taking the reach of rumors into the economic domain
\cite{strauss2013sec}. In 2016 during the politically divisive Brexit and US elections, fake news outpaced real news on Facebook \cite{Sil16}. Note that we use ``rumor" and ``fake news" interchangeably in this work, as is common in related work.

Given these very real social and economic implications of rumors in social media, automatic detection of rumors has seen a significant surge in research. Existing work in this area use some aspects of the news item like (i) news item text content, (ii) 
comments on the news item (iii) user characteristics and (iv) 
propagation paths of the item within the network.  Some 
researchers have tackled this problem by creating knowledge graphs that are built by crawling the web for raw facts \cite{PawPalBha17} and then further processing and cleaning it up and using it to fact check \cite{CiaEtal15}. The problem with this approach is that it is less suitable for detecting rumors in evolving content that don't yet have a representation in the knowledge graph.
Among other methods that have used the content of the news item, approaches have ranged from using psycholinguistic features like sentiment \cite{CasMenPob11}, style features like readability \cite{PerEtal18} and assertive and factive verbs \cite{Pop17} with varying degrees of success. 

Some authors have suggested that other characteristics may be
useful to include in the detection process due to the insufficiency of the news content material, especially in microblogging sites like 
Twitter \cite{LiuWu18}. 
Several researchers have introduced other information like user comments \cite{MaEtal2016} along with news content; user characteristics like number of followers, the first to tweet a story etc \cite{LiEtal19}. Still others have used the network structure and/or propagation path along with content \cite{LiuWu18,MaEtal2017}.

A hybrid feature extraction unit (HFEU) and a gated diffusive unit (GDU) were used to detect rumors in \cite{ZhaEtal20}. 
HFEU extracted explicit and latent features from the textual information; GDU effectively extracted relationship among news articles, creators and subjects.
A fake news detector called event adversarial neural networks (EANN) that includes a multi-modal feature extractor, a fake news detector and an event discriminator which co-operatively learns event non-specific features to discriminate between fake and real news was proposed in \cite{WanEtal18}. 

More recently, authors of \cite{MohEtal18} argued 
that interpretable news feed generator algorithms 
could 
reduce their misuse by improving user awareness and system transparency. 
T-SNE based methods were provided in \cite{MonEtal19} which could indicate the usefulness of learned features for rumor classification.
Research has now begun in explainable rumor detection algorithms \cite{khoEtal20}. 

Another debate that is often waged in the AI community, is whether handcrafted 
features should/can be incorporated in the AI engine. In this work we
provide a framework that can be used to explore this question by 
including both handcrafted 
and latent features for the rumor detection problem.

\emph{We  propose a modular, explainable architecture that can use any number of classes of features that may become available, for detecting rumors.}
Specifically, we design an explainable deep learning architecture 
using attention mechanism to detect 
rumors using multiple types of features.
Our work is inspired by \cite{khoEtal20} but with some differences and can be thought of as a generalization to multiple class of features.
The main contributions of our work are:
\begin{itemize} \itemsep 2pt
     \item modular architecture that can be extended to as many feature classes as desired
     \item inter and intra-feature attentions that capture the relative importance between the different feature classes as well as the relative importance of features within a class. These can be used to provide explanation of the model's conclusions
    \item extensive testing on popular datasets and insights into the use of latent and handcrafted features 

\end{itemize}

\section{Related Work}
\label{sec:related work}
One way to classify existing fake news and rumor detection approaches, is to think of them from the perspective of (i) knowledge, (ii) style, 
(iii) propagation and (iv) credibility \cite{ZhoEtal19}. Knowledge based methods essentially use knowledge bases to verify any given news item like the B-TransE model based on news content using knowledge graphs is proposed in \cite{PanEtal18}.

Style-based fake news detection posits that there are stylistic
differences between fake and real
news. These differences are quantified and used to differentiate between fake and real news. Typically psycholingusitic and linguistic features are used in this category of work. Examples of these features 
include special characters, emotion symbols, sentiment (positive/negative) words, hashtags \cite{CasMenPob11}, part-of-speech tags and lexicon patterns \cite{QazEtal11}, swear words and pronouns
\cite{GupEtal14}, etc.

Propagation-based fake news detection uses information found in news dissemination. Some researchers have used propagation paths to detect rumors \cite{MaEtal2017,LiuWu18}, while others have extracted temporal-linguistic
features from user comments \cite{ZheEtal2015,MaEtal2016}. An approach to detecting fake news was proposed by creating a feature vector from user characteristics and model five minutes of tweets as a time series and using this propagation based feature with a CNN+RNN model to detect fake news \cite{LiuWu18}.

Credibility-based fake news detection algorithms assesses the credibility of headlines by using various methods like click-bait
detection \cite{ShuEtal17}, publisher information, comments \cite{MaEtal2017,ZheEtal2015,MaEtal2016}, and user characteristics \cite{CasMenPob11}. 
Some researchers use style to determine credibility \cite{GupEtal14}. 
Profile features like number of followers, number of friends and registration age have also been used to detect rumors \cite{CasMenPob11}.
Some more recent methods don't truly fall into the above classification scheme, like the work in
\cite{GuoEtal18} where a hierarchical bidirectional LSTM model is used for representation learning and social contexts are incorporated into the network via attention mechanism to detect rumors.
Some researchers use the relationship between user and content to detect rumors 
with a combined model based on deep learning, which includes convolutional neural network
(CNN) and long short-term memory (LSTM)\cite{ZhouEtal20}. The sentiment in the comment 
was also used as an important feature in a CNN-LSTM model to detect rumors in \cite{LvEtal20}.  

Work on explainable AI, in general, has started to emerge specifically to address the problem of trustablity of AI systems \cite{GilEtal18,ZhaEtal20}.
Within the context of explainable fake news detection, an algorithm called XFAKE was developed in \cite{YanEtal19} where
attributes (e.g., speaker, context, etc.) and statements are analyzed.
It proposes three frameworks for fake news detection using word2vec \cite{MiEtal13}: (1) MIMIC which combines deep neural network to learn and a shallow model to interpret;
(2) ATTN framework which consists of self-attention mechanism and convolutional neural network; 
(3) PERT framework which uses XGBoost together with perturbation-based method. However, 
these methods either use drop out methods or train an ensemble model first and then 
distill it to a smaller model (XGBoost) in order to explain the importance of features. 
A multi-source multi-class fake news detection framework (MMFD) was proposed in \cite{KarEtal18}, which is an interpretable method to integrate information from multiple sources potentially providing information with various degrees of fakeness \cite{WangEtal17}. 
It uses attention mechanism to identify contribution of each source.

An explainable fake news detection (dEFEND) model was developed in \cite{KaiEtal19} to 
extract the top $k$ check-worthy sentences and user comments for fake news detection. 
Their work treats the tweet and comments separately and uses co-attention
mechanism to identify the top $k$ important sentences and comments. 

Instead of adopting popular tree-based model, a flattened-tree
structure was proposed in \cite{khoEtal20}, which arranged all tweets in a chronological order. A post-level attention model (PLAN), a structure aware self-attention model (StA-PLAN) and 
a hierarchical token and post-level attention model (StA-HiTPLAN) were 
proposed to explain both post-level and token-level rumor detection 
predictions. They modify the encoder structure of the transformer 
\cite{vasEtal17} and do away with heavier neural networks in their architecture.
Our architecture is inspired by their work but is different in several ways: 
(1) we propose a mechanism that allows for the addition of several classes 
of features, including handcrafted features, thereby allowing for a
``best-of-both-worlds" type approach; (2) we use attention mechanisms to
provide explanation of the AI model in terms of the most important
inter-class and intra-class features (3) we use positional encoding \cite{vasEtal17} to 
maintain the relative positions of the tweets and comments in a conversation, whereas time
delays was used in \cite{khoEtal20} (4) our architecture is modular
in that, other classes of features can be added on if needed (5) we use sentence embedding instead of word embeddings and (6) we can provide explanations from general to specific \cite{YangEtal19}. 
We demonstrate
this approach using
latent features derived from universal sentence embeddings of the tweets and their
responses; features derived from user information as well as handcrafted
features from the content. 
Experiments on popular datasets show that our architecture outperforms
the work in \cite{khoEtal20} as well as other benchmark algorithms.

\section{Model and Method}
\label{sec:model and method}
We propose an explainable AI model for rumor detection
using attention mechanisms \cite{vasEtal17} to interpret the decisions that 
the AI models make. Our model is modular in that it can accommodate different
classes of features as needed. In this work we illustrate the model with
three types of features. We use inter-feature and intra-feature attention
to explain the relative importance of the features in the final 
outcome. This general architecture can accommodate handcrafted as well as 
latent features and provide explanations. 

We use the universal sentence embedding (USE) \cite{CerEtal18} to capture 
the latent features of the contents of the microblog posts as well as its 
responses. The USE is used to provide a vector representation of each of the 
tweets and their responses. 
We also use two classes of handcrafted features: one based on the users' 
profile and the other on the content of the microblog post and its responses.
\begin{figure}[hbt!] 
\centering
\includegraphics[width=0.48\textwidth]{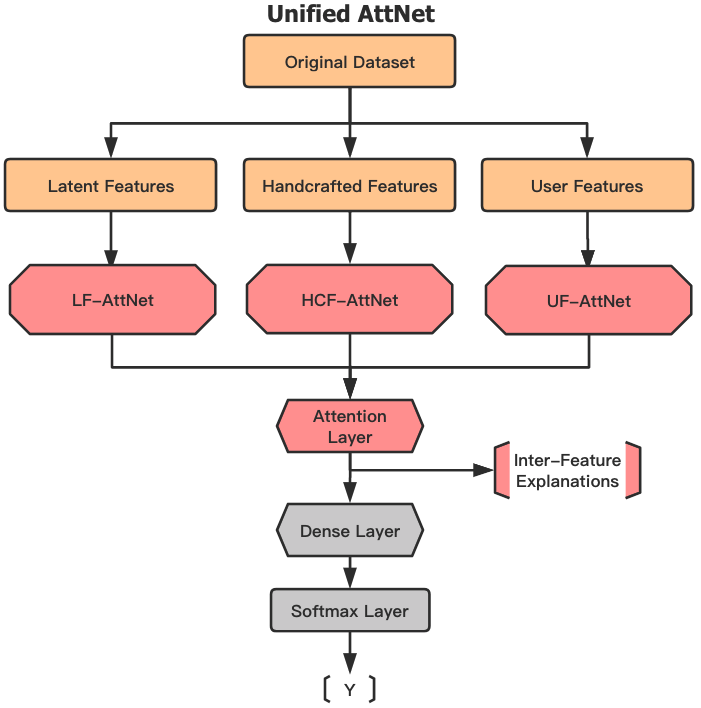}
\caption{The proposed modular, unified attention network for three classes of features: latent feature derived from content of the post; handcrafted features derived from the content and responses and user profile derived features. Note that more ``legs" can be added if desired to capture other feature classes.}
\label{fig:unified}
\end{figure}

The handcrafted content level features include: parts of speech (PoS) tags, 
vocabulary richness measures, sentiment, readability and linguistic features.
The use of vocabulary richness and readability score features
is motivated by work done in fake news that shows that there is a strong correlation between the language proficiency exhibited in 
the 
item and whether it is true or not \cite{Farm19}.
Researchers have also shown that
fake news often tends to be dramatic and emotional in content \cite{GuoEtal19, AnoEtal19}, 
and that user features can
play a role in detecting rumors \cite{CasMenPob11,ChaEtal16,LiEtal19}. So we use both sentiments and user features in the proposed method.

\begin{figure}[hbt!] 
\centering
\includegraphics[width=0.48\textwidth]{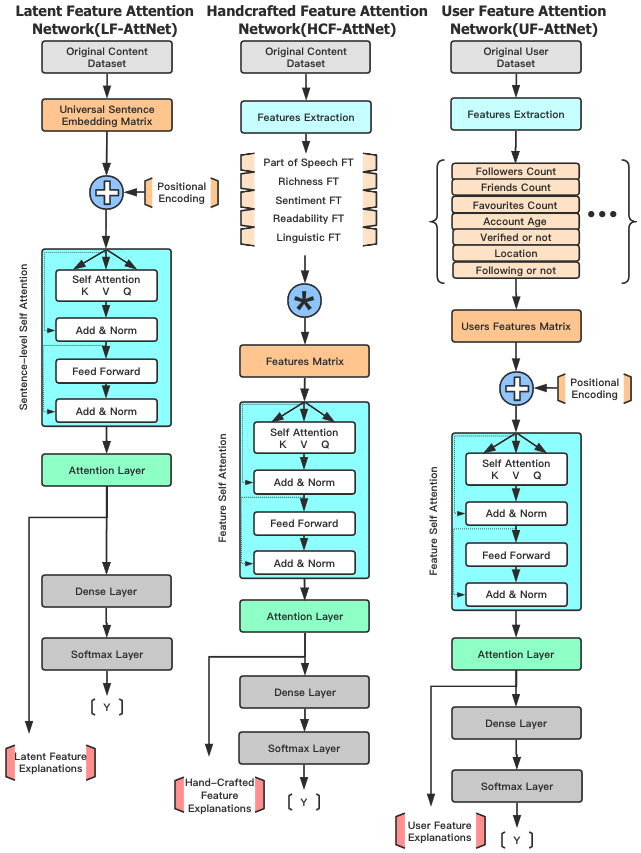}
\caption{The proposed architecture of Latent Feature Attention Network(LF-AttNet), Handcrafted Feature Attention Network(HCF-AttNet) and User Feature Attention Network(UF-AttNet). LF-AttNet uses the sentence embeddings of the tweet, HCF-AttNet uses handcrafted features extracted from content and UF-AttNet uses user features.}
\label{fig:three_legs}
\end{figure}

Fig.~\ref{fig:unified} shows the proposed architecture and comprises of three legs, which we call (1) the Latent Feature Attention 
Network (LF-AttNet), (2) Handcrafted Feature Attention Network (HCF-AttNet) and 
(3) User Feature Attention Network (UF-AttNet). These three legs are combined 
using another attention layer followed by a dense layer and the softmax layer.
The dense linear layer is the same as that 
proposed in the transformer \cite{vasEtal17}. 
The attention layer captures the relative 
importance among content, users' and the USE features and helps provide an interpretation at 
the feature class level.
The three
``legs" of this architecture are shown in Fig~\ref{fig:three_legs} and described in the following subsections. 

\subsection{Latent Feature Attention Network}
\label{sec:emb-net}
The architecture of the proposed Latent Feature Attention Network(LF-AttNet) is the leftmost 
subfigure in Figure \ref{fig:three_legs}. We propose this architecture as a means of capturing latent feature information implicitly in language embeddings 
via the universal sentence 
embedding (USE) \cite{CerEtal18} of
each tweet and response.
A positional encoding module is 
used to maintain the relative positions of
the tweet and its responses. 
This approach was used in the 
transformer 
\cite{vasEtal17} architecture to 
keep the relative positions
of the sentences intact. This is different from the work in \cite{khoEtal20} where a time delay encoder is used.

Let $u_i$ be the USE vector corresponding to the $i^{\rm {th}}$ 
tweet in the record, where $i$ ranges through all tweets and their responses.
Positional encoding is applied to each 
tweet and response and the resulting vectors,  $u'_i$, are used to construct the matrix
$U=\{u'_1,u'_2,...,u'_n\}$. An $h$-layer multi head attention
(MHA) module is used to
extract the intra-feature relationships in this architecture.
This MHA module uses the scaled dot product attention 
\cite{vasEtal17}, which is given by 
\begin{equation}
{
  {\rm Attention}(Q,K,V)= {\rm softmax} \left (\frac{QK^T}{\sqrt{{d}_k}} \right)V}
 \end{equation}
 where $Q,K$ and $V$ are the query, key and value matrices and $d_k$ is the dimension of the query and key vectors.
This is followed by an attention layer that captures
interpretation at the latent feature level. 
The output of the attention layer is fed to a 
linear layer and a softmax layer to get the final 
prediction.

\subsection{Handcrafted Feature Attention Network}
\label{sec:FT-net}
The architecture is shown in the 
middle subfigure of Figure~\ref{fig:three_legs}. 
This architecture (HCF-AttNet) also comprises of
a self-attention module that captures 
the intra-feature relationships; an attention layer that can be used 
to generate feature level explanations followed by a linear layer and 
a softmax layer.
The content of the tweet and responses are analyzed and the following features are extracted: for each record, $27$ PoS tags and $1$ sentiment feature(average value of words' sentiment polarity) by using NLTK \cite{LopEtal02}, $4$ psycho-linguistic features (`FamiliarityScore', `ConcretenessScore', `ImagabilityScore' and `AgeofAcquisitionScore'), 
$4$ vocabulary richness features (Honore’s Statistic (HS), Sichel Measure (SICH), Brunet’s  Measure(BM) and Text-Type Ratio (TTR)) and $2$ readability features (Automated ReadabilityIndex (ARI) and Flesch-Kincaid readability (FKR) scores) \cite{FraEtal16}.
Examples of some of these PoS tags are shown in Table~\ref{tab:postags}.
The handcrafted features for the $i^{\rm {th}}$ tweet (or response) are 
gathered into a vector $c_i$, which is then arranged into a matrix $C=\{c_1,c_2,...,c_n\}$.
An $h$-layer Multi-Head-Attention (MHA) module is applied
to $C=\{c_1,c_2,...,c_n\}$ to capture the relationship between the handcrafted features.
The MHA module is followed by a linear dense layer \cite{vasEtal17} and softmax layer to get the final classification.
\begin{table}[ht]
\begin{center}
\begin{tabular}{p{1.4cm}c} 
\hline
\hline
\textbf{Tag} & \textbf{Description}  \\ 
\hline
\hline
SYM&Symbol\\
RB&Adverb\\
CD&Cardinal number\\
JJ&Adjective\\
VBZ&Verb,3rd person singular present\\
MD&Modal\\
PRP&Personal pronoun\\
NNP&Proper noun, singular\\
NNS&Noun, plural\\
WRB&Wh-adverb\\
CC&Coordinating conjunction\\
VBG&Verb, gerund or present participle\\
VB&Verb, base form\\
PDT&Predeterminer\\
VBD&Verb, past tense\\
...& ...\\
\hline
\hline
\end{tabular}
\caption{Examples of PoS tags feature}
\label{tab:postags}
\end{center}
\end{table}
\subsection{User Feature Attention Network}
\label{sec:user-net}
As mentioned earlier, researchers have shown that user profile features may also be indicative of whether
or not a tweet is fake \cite{CasMenPob11,ChaEtal16,LiEtal19}. 
The UF-AttNet architecture is proposed on the right side of Figure \ref{fig:three_legs}. Particularly features like: ``followers count", ``friends count", ``favorites count" (how many favorites the user has), ``account age"
(how long the account was active before the tweet was sent), 
``verified or not" (whether or not the user is verified), 
``location" (geographic location of the user's), ``following or not" (whether this user follows other users or not).

We propose to use these features as well and group them
under the user feature category. These features
are extracted and converted to a vector. 
Position encoding is applied to these vectors to 
ensure that the user characteristics of the original
poster and all responses are kept distinct.
The rest of the architecture is the same as the 
proposed HCF-AttNet and is depicted in the rightmost subfigure of Fig.~\ref{fig:three_legs}.

\section{Experiment and Result}
\label{sec:experiment}

We use the PHEME \cite{ZubEtal15, ZubEtal16} and RumourEval2019 \cite{GorEtal18} datasets to 
evalute our proposed architecture and its variants.
The PHEME dataset is
a collection of Twitter rumors and non-rumors posted during five breaking news events including, Charlie Hebdo (458 rumors (22.0\%) and 1,621 non-rumors (78.0\%)); Ferguson (284 rumors (24.8\%) and 859 non-rumors (75.2\%)); Germanwings Crash (238 rumors (50.7\%) and 231 non-rumors (49.3\%)); Ottawa Shooting (470 rumors (52.8\%) and 420 non-rumors (47.2\%)) and Sydney Siege (Sydney Siege: 522 rumors (42.8\%) and 699 non-rumors (57.2\%)).
The dataset also contains the tweet IDs and responses to tweets.
Rumors are annotated for their veracity as ‘true’(T, for true rumor), ‘false’(F, for false rumor i.e fact) or ‘unverified’(U).
The RumourEval2019 dataset is a collection of rumors from Twitter and Reddit, including 381 twitter threads and 65 Reddit threads. 
The dataset is annotated using ‘true’(T), ‘false’(F) or ‘unverified’(U); however this dataset does not include user features.

\subsection{Experimental Setup}
\label{sec:setup}
We implemented our proposed model in Pytorch and 
trained it to minimize the cross-entropy 
loss function of predicting the class label of tweets in the training set. As mentioned 
earlier, we extracted three categories of features:
content linguistic features \cite{FraEtal16}, user 
features and latent features derived from the USE
embedding of tweets and responses.  
For all models in our 
experiments, we used a $6$-layer multi-head attention (MHA) module and the stochastic 
gradient descent + momentum (SGD + Momentum) \cite{Rud16} as the optimizer for training.
Since conversations (tweets and their responses) can be 
of varying lengths, we used the average number of tweets
and responses as the conversation length for the dataset.
We truncated conversations that were longer than the 
average length and padded 
with empty strings for those with shorter conversations. 
Note that changing this number to the median number of tweets or the maximum number of tweets did not give us significantly different results. 
We split the original data in the $4:1$ ratio for
training and testing and used $5$-fold cross validation to prevent over-fitting the model. 
Note that this was the same experimental set-up used in the benchmark 
algorithms that we compare against.

\begin{table}[hbt!]
\centering
\begin{tabular}{ccc} 
\hline
\hline
\textbf{Method}  &\textbf{F1 Score}&\textbf{Accuracy}    \\  
\hline
\textbf{Majority (True)} & 0.226 & 0.511 \\
\textbf{NileTMRG} & 0.339 & 0.438 \\
\textbf{BranchLSTM} & 0.336 & 0.454 \\
\textbf{MTL2} & 0.376 & 0.441 \\
\textbf{MTL-SL} & 0.418 & 0.483 \\
\textbf{PLAN} &0.360&-\\
\textbf{StA-PLAN}& 0.349&-\\
\textbf{StA-HiTPLAN} &0.379&-\\
\textbf{PLAN+TD} &0.386&-\\
\textbf{StA-PLAN+TD} &0.369&-\\
\textbf{StA-HiTPLAN+TD} &0.395&-\\
\hline
\textbf{LF-AttNet} & 0.437& \textbf{0.559}\\
\textbf {HCF-AttNet}&\textbf{0.453}&0.555\\
\textbf{UF-AttNet}&0.430&0.539\\
\textbf{Unified-AttNet}& 0.448&0.551\\
\hline
\hline
\end{tabular}
\caption{The comparison of performance of rumor verification between our models and benchmarks for the PHEME dataset (The experiment setup is same): Majority (True), NileTMRG, BranchLSTM, MTL2 and MTL-U(Multi Task Learning-User Info) were referenced from \cite{LiEtal19}, all these methods use the users' information for prediction. PLAN, StA-PLAN, StA-HiTPLAN, PLAN+TD, StA-PLAN+TD and StA-HiTPLAN+TD were referenced from \cite{khoEtal20}. The proposed
methods appear below the horizontal line. UF-AttNet uses users' features, the HCF-AttNet uses handcrafted content based features, the LF-AttNet uses sentence embeddings and the Unified-AttNet is a combination of LF-AttNet, HCT-AttNet and UF-AttNet.}
\label{tab:com_verification}
\end{table}
\begin{table}[hbt!]
\centering
\begin{tabular}{ccc} 
\hline
\hline
\textbf{Method}  &\textbf{F1 Score} &\textbf{Accuracy}  \\  
\hline
\textbf{BranchLSTM} & 0.3364&- \\
\textbf{NileTMRG} & 0.3089&-\\
\hline
\textbf{LF-AttNet} & 0.3414&0.4033\\
\textbf {HCF-AttNet}& \textbf{0.4148}&\textbf{0.5020}\\
\textbf{LF+HCF-AttNet}& 0.3715&0.4603\\
\hline
\hline
\end{tabular}
\caption{The comparison of performance of rumor verification between our models and benchmarks for the RumourEval2019 dataset (the experiment setup is same): BranchLSTM and NileTMRG were referenced from \cite{LiEtal19}. LF+HCF-AttNet is a combination of LF-AttNet and HCT-AttNet. Since user information isn't available in this dataset our unified architecture consists of only two legs. This is an example of the modularity of this architecture.}
\label{tab:com_verification_19}
\end{table}

\subsection{Benchmarks}
\label{sec:results}
We compare the performance of our architectures 
against 11 benchmark algorithms and two datasets: PHEME and RumourEval19
in terms of accuracy and F1 score. 
These comparisons are shown in Table~\ref{tab:com_verification} and 
Table~\ref{tab:com_verification_19} respectively. 
The benchmark architectures are described below.
\begin{itemize}
\itemsep -2pt
\item \textbf{Majority (True):} A strong baseline which results in high accuracy due to the class imbalance in the veracity classification task \cite{Jam98}.
\item \textbf{NileTMRG:} A veracity prediction system from SemEval-2017 Task 8 \cite{EnaEtal17}, based on a linear SVM using a bag-of-words representation of the tweet concatenated with selected features.
\item \textbf{BranchLSTM:} A method based on an LSTM layer followed by several dense ReLU layers and a softmax layer \cite{ZubEtal18}.
\item \textbf{MTL2:} A multi-task learning method without task specific layers \cite{KocEtal18}.
\item \textbf{MTL-SL:} A multi-task learning method with shared layer \cite{LiEtal19}.
\item \textbf{PLAN:} A post-level attention network with GLOVE 300d embedding \cite{khoEtal20}.
\item \textbf{StA-PLAN:} A structure aware post-level attention network with GLOVE 300d embedding \cite{khoEtal20}.
\item \textbf{StA-HiTPLAN:} A structure aware hierarchical token and post-level attention network with GLOVE 300d embedding \cite{khoEtal20}.
\item \textbf{PLAN+TD:} PLAN model with time delay information \cite{khoEtal20}. 
\item \textbf{StA-PLAN+TD:} StA-PLAN model with time delay information \cite{khoEtal20}.
\item \textbf{StA-HiTPLAN+TD:} StA-HiTPLAN model with time delay information \cite{khoEtal20}.
\end{itemize}

\subsection{Performance Analysis}
\label{sec:perf-anal}
In this section we will analyze the performance of the proposed architectures in terms of the
traditional metrics including F-score and accuracy as well as in terms of explainability.

From Table~\ref{tab:com_verification}, we see that all the proposed architectures perform better
than the benchmark architectures on the PHEME dataset using the same experiment setup as the benchmark algorithms.
The proposed HCF-AttNet
performs best, the Unified-AttNet second best and LF-AttNet and UF-AttNet performing very close to each other
in terms of F1 score. The accuracy of all proposed architectures are
close to each other and significantly better than the other benchmarks.
Since F1 scores tend to provide a balance between precision and recall and when the data 
is imbalanced, this may be a more appropriate metric to 
assess the performance of a rumor detection algorithm.
Note also that the proposed architectures perform better than
the explainable rumor detection architecture proposed in \cite{khoEtal20}. 
It would also appear that
on an average the handcrafted features seem to be doing better than only latent features
extracted from the content of the tweets and their responses.


From Table~\ref{tab:com_verification_19}, all the proposed models outperform the two baseline ones using the same experiment setup as the comparison models; with the HCF-AttNet performing best. 
We notice that the trend seems to be in favor of HCF-AttNet across the different types of datasets.
\emph{This points to the fact that handcrafted features can play an important role in the 
detection of rumors.}

\subsection{Explainability Analysis}
In this section we analyze the explainabilty of 
the algorithms using PHEME and RumourEval2019 
dataset. Before we continue with the rest of this subsection, we define ``influence scores'' which 
we will use to measure the influence of various types of features on the final decision.
Note that since the dataset has three categories of data our model is a three class classifier
that outputs one of the three labels T, F or U corresponding to whether the model classifies this as a true rumor (T, essentially a rumor); false rumor (F, essentially a fact) or unverified (U).
\subsubsection{Influence Score}
In order to understand the contribution of the different classes of features (or legs of the 
architecture) we define a metric called 
\textbf{inter-class influence score} as described below. 
Let the attention value for a particular decision, for each leg in Fig.~\ref{fig:three_legs} 
be $\alpha^{l}$, $\alpha^{h}$ and $\alpha^{u}$ for the
latent feature leg, the handcrafted feature leg and the user feature leg, respectively. 
Define $\mathbf{\alpha}_i$ as a vector: $\mathbf{\alpha}_i=[\alpha^l_i,\alpha^h_i,\alpha^u_i]$, where $i$ represents the $i^{\rm th}$ data record. 
Let $\mathrm A =[{\mathbf{\alpha}}_0,\mathbf{\alpha}_1,\mathbf{\alpha}_2,...,\mathbf{\alpha}_i,...]$ be
a matrix of attention vectors $\alpha_i$. 
We can define an inter-class influence score for each decision using these attention values.
For each correct decision, we count the number of times 
the attention score for the latent features is highest
(compared to all three attention scores) and define  the influence score ${\rm Influ}_{l}^c$ for the latent feature leg as the frequency of the event that the attention value corresponding to the latent feature leg was the highest.
Similarly we can define the influence scores, ${\rm Influ}_{h}^c$ and ${\rm Influ}_{u}^{c}$ for the other two legs.  That is,
${\rm Influ_{l}^{\rm c}}= \frac{N^{\rm c}_{l}}{N^{\rm c}_{\rm data}}$,
${\rm Influ_{h}^{\rm c}}= \frac{N^{\rm c}_{h}}{N^{\rm c}_{\rm data}}$,
${\rm Influ_{u}^{\rm c}}= \frac{N^{\rm c}_{u}}{N^{\rm c}_{\rm data}}$,
 where $N^{c}_{\rm data}$ is the total number of data records where the correct decision was made
 and $N^c_{l}$, $N^c_{u}$ and $N^c_{h}$ are the number
 of times that the latent feature attention weights, the user feature attention weights and the handcrafted feature attention weights were the highest, respectively.
 
Similarly, we also define the influence scores for the three legs when the decisions were incorrect. Note that we can think of  incorrect decisions in three ways: when a the classifier gives a true rumor the wrong label: false rumor (fact) or unverified:``T $\rightarrow$ F/U'', when a false rumor is given the wrong label (``F $\rightarrow$ T/U'')  and when an unverified piece of information is given the wrong label (``U $\rightarrow$ T/F''). 
We can calculate the influence of each of these legs for each kind of incorrect decisions as follows. Let $N^{\rm inc\_T}_i$ represent
the number of times the attention weight for the $i^{\rm th}$ feature leg was the highest when the true rumor is mislabled; where
$i = \{l,u,h\}$ represent the different classes of features as before. Let $N^{\rm inc}_{\rm data}$ be the number of data points 
for which the decision was incorrect.
Then 
for ``T $\rightarrow$ F/U'', the influence scores for the three
legs can be 
written as
${\rm Influ}_{l}^{\rm inc\_T} = \frac{N^{\rm inc\_T}_l}{N^{\rm inc}_{\rm data}}$, ${\rm Influ}_{h}^{\rm inc\_T} = \frac{N^{\rm inc\_T}_h}{N^{\rm inc}_{\rm data}}$, ${\rm Influ}_{u}^{\rm inc\_T} = \frac{N^{\rm inc\_T}_u}{N^{\rm inc}_{\rm data}}$. 

Similarly, when false rumor (or a fact) is mislabeled, ``F $\rightarrow$ T/U'', we can compute the influece scores as
${\rm Influ}_{l}^{\rm inc\_F} = \frac{N^{\rm inc\_F}_l}{N^{\rm inc}_{\rm data}}$, ${\rm Influ}_{h}^{\rm inc\_F} = \frac{N^{\rm inc\_F}_h}{N^{\rm inc}_{\rm data}}$, ${\rm Influ}_{u}^{\rm inc\_F} = \frac{N^{\rm inc\_F}_u}{N^{\rm inc}_{\rm data}}$.
Finally when an unverified data point is misclassified,
``U $\rightarrow$ T/F'', we have:
${\rm Influ}_{l}^{\rm inc\_U} = \frac{N^{\rm inc\_U}_l}{N^{\rm inc}_{\rm data}}$, ${\rm Influ}_{h}^{\rm inc\_U} = \frac{N^{\rm inc\_U}_h}{N^{\rm inc}_{\rm data}}$, ${\rm Influ}_{u}^{\rm inc\_U} = \frac{N^{\rm inc\_U}_u}{N^{\rm inc}_{\rm data}}$.
We can also define influence scores for each feature
inside each class of features (intra-class influence scores) in a similar way (formal definitions not provided for brevity reasons).
\subsubsection{Average Case Analysis}
\begin{figure}[!htb]
	\begin{center}
		\subfigure[]{\includegraphics[width=10cm]{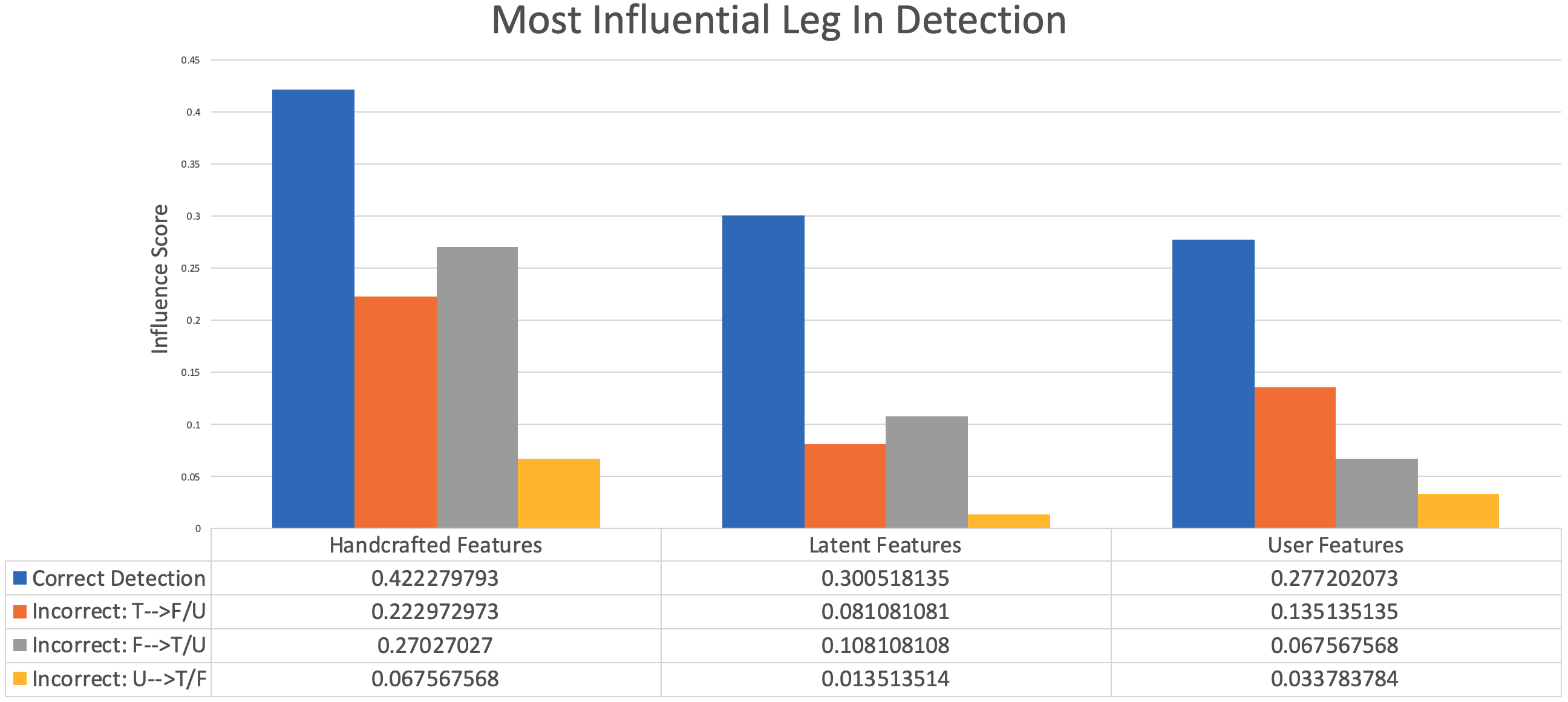}}\hfill
		\subfigure[]{\includegraphics[width=10cm]{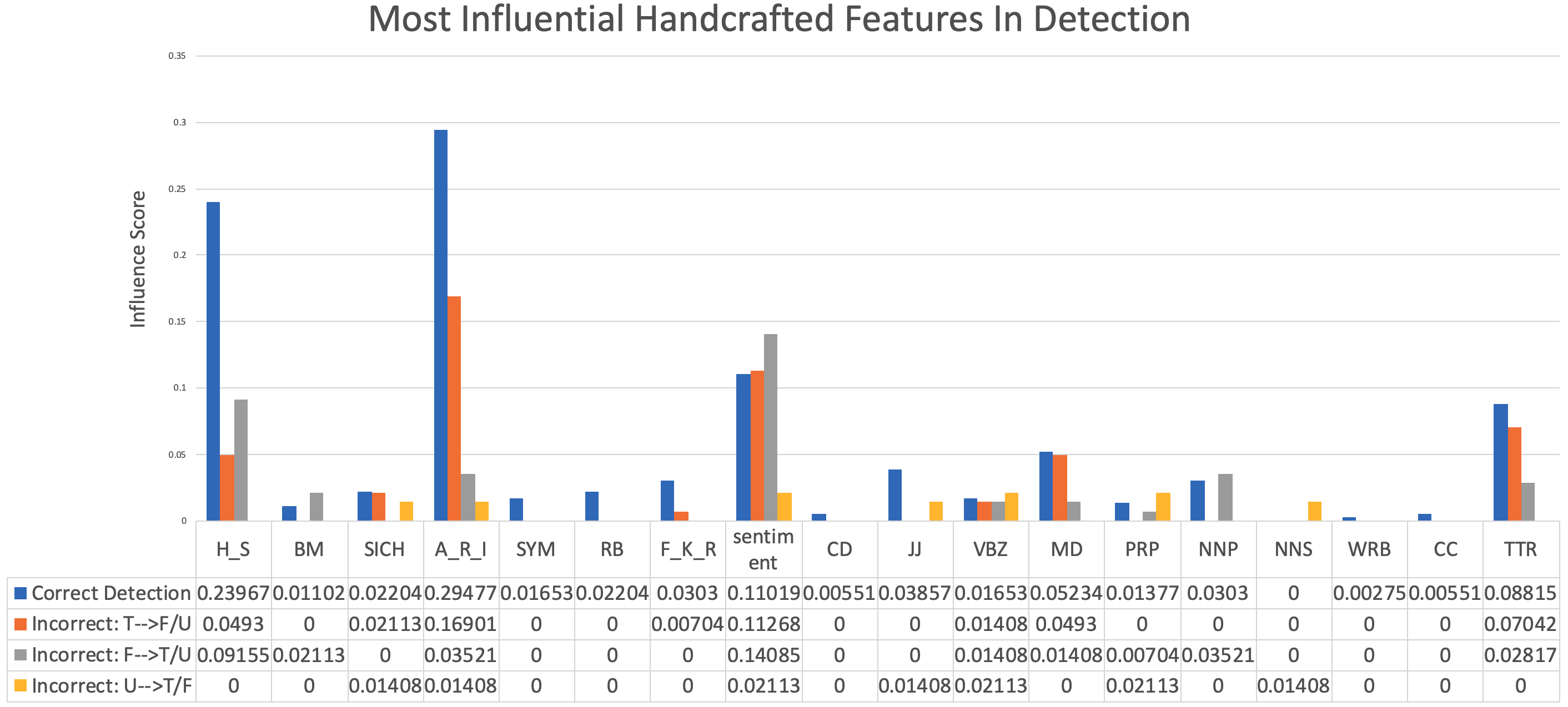}}\hfill
		\subfigure[]{\includegraphics[width=10cm]{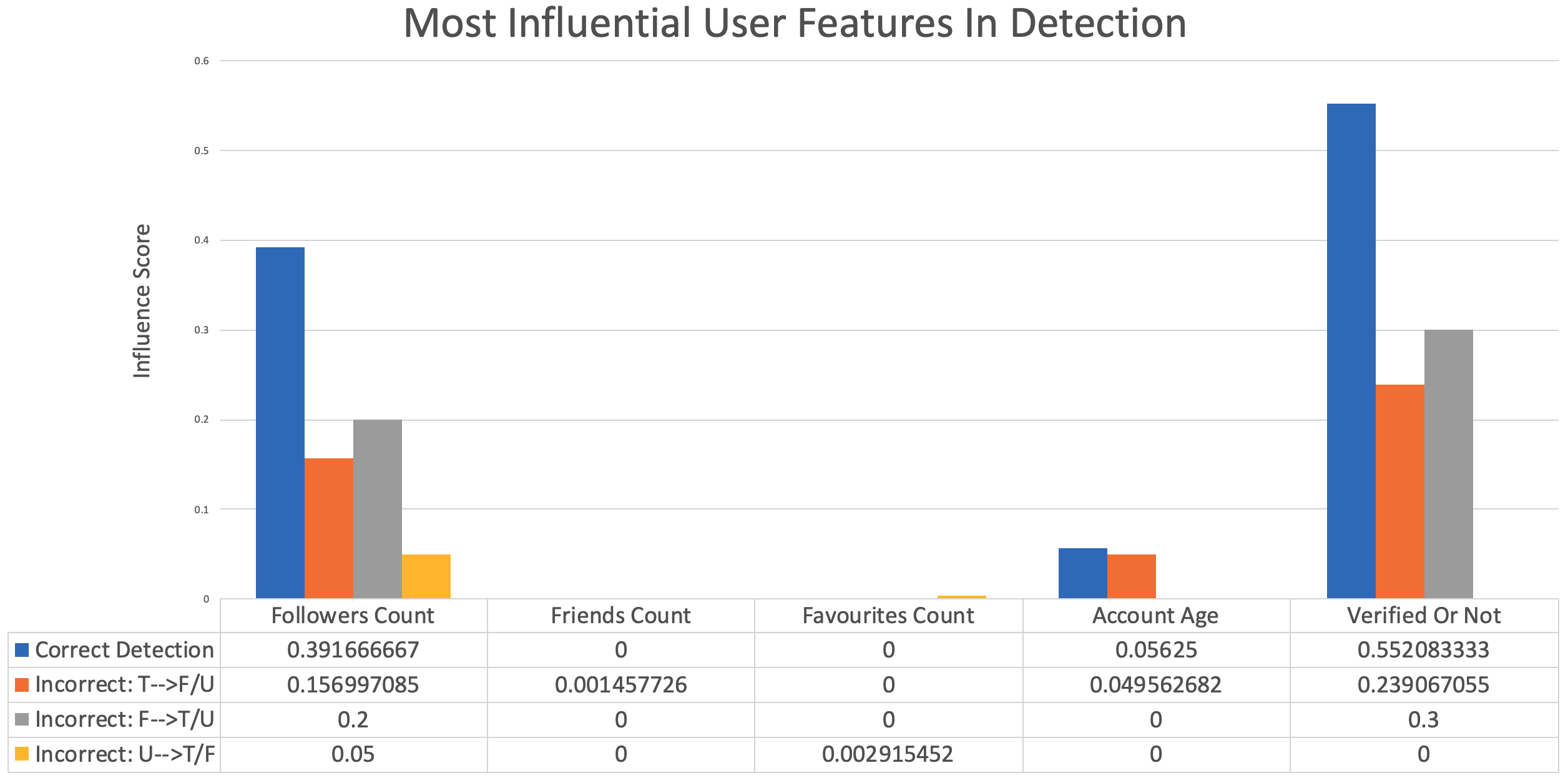}}\hfill	
	\end{center}
	\caption{Influence score histograms on PHEME dataset. (a) Histogram of influence scores for each of the three legs when the correct (blue) and incorrect decisions are made. In incorrect decisions: ``T $\rightarrow$ F/U'' means the true rumor is misclassified; (b) histogram of the most influential handcrafted features and (c) histogram of most influential user features.}
\label{fig:influ}
\end{figure}

\begin{figure}[!htb]
	\begin{center}
		\subfigure[]{\includegraphics[width=10cm]{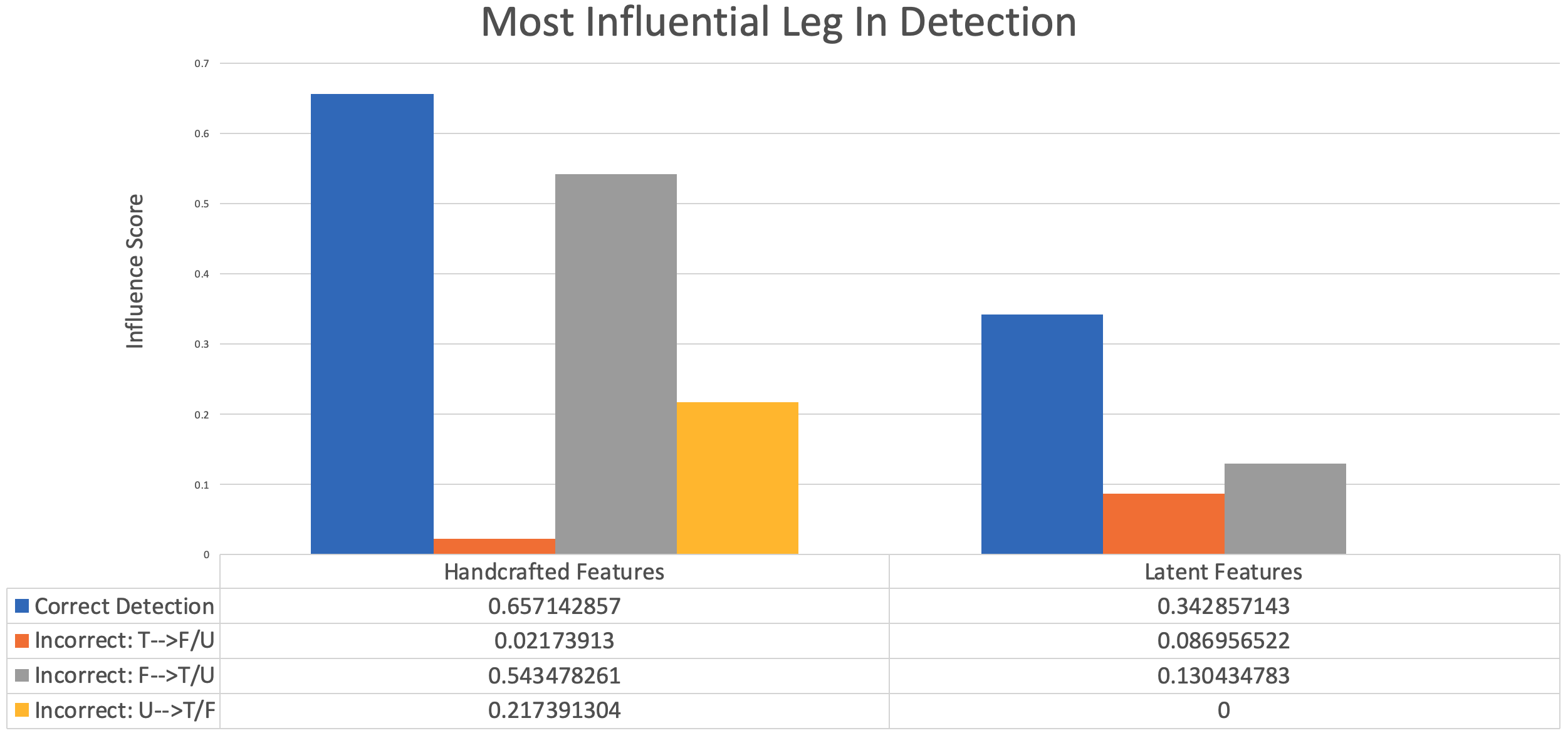}}\hfill
		\subfigure[]{\includegraphics[width=10cm]{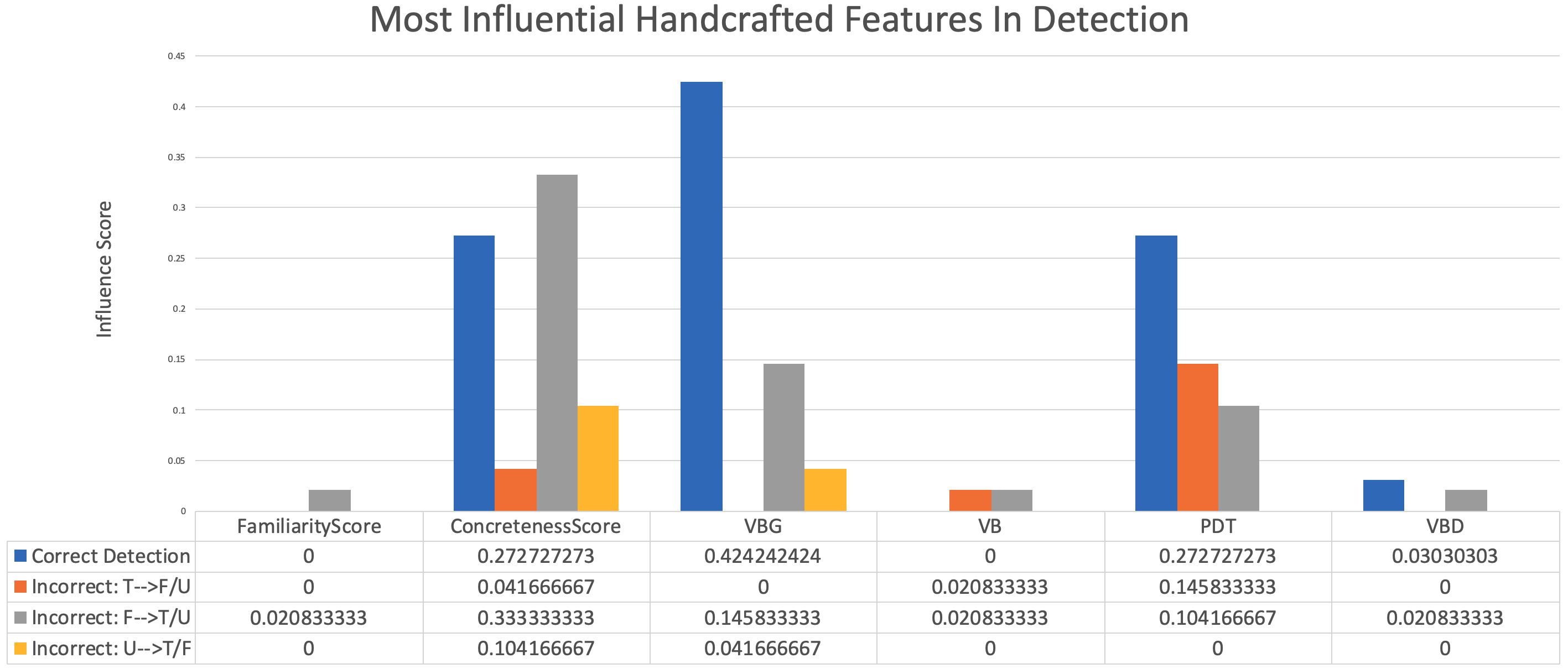}}\hfill
	\end{center}
	\caption{Influence score histograms on RumourEval2019 dataset. (a) Histogram of influence scores for each of the three legs when the correct (blue) and incorrect decisions are made. In incorrect decisions: ``T $\rightarrow$ F/U'' means the true rumor is misclassified; (b) histogram of the most influential handcrafted features.}
\label{fig:influ_new}
\end{figure}

Fig.~\ref{fig:influ}a 
shows the inter-feature influence scores for each of the three legs in the unified architecture for
both incorrect and correct decisions using the PHEME dataset. For both correct and incorrect detection, handcrafted features 
contributes most. This implies that handcrafted features play an important role in the outcome. 
However, the latent features and user features together account for $58\%$ of the self attention values in
correct detection and $44\%$ in incorrect detection, which means the latent and user features also play significant roles and cannot be ignored. 

Fig.~\ref{fig:influ}b shows the influence score for handcrafted features.
Feature H\_S (Honore’s Statistic), A\_R\_I (Automated Readability Index), sentiment and TTR (Text-Type Ratio) together
account for $69\%$ when the decision is correct and $73\%$ in 
incorrect decisions. This indicates that these four features contain more information than others on an average. 
H\_S and A\_R\_I contribute more to 
correct detection than incorrect decisions, whereas sentiment 
contributes more to incorrect detection. The influence scores of TTR for correct detection and incorrect detection are close.
These observations are in accordance with some of the prior studies 
on the sensational nature of fake tweets, language sophistication and
clarity of language
and the role of these characteristics in fake news detection \cite{LiuWu18}. 

Fig.~\ref{fig:influ}c shows the influence score for user features. 
From this figure we see that whether the account is verified or not is an
important indicator of
whether the tweet is trustworthy or not and so is the followers count feature. Although the age of the account contributes somewhat,
it does not seem to have as big an influence. This is true for both the correct and incorrect decisions. This would suggest that we may be able to drop all but these three features in the model going forward.

Fig.~\ref{fig:influ_new}a 
shows the influence scores for each of the two legs in the unified architecture for
both incorrect and correct decisions. Like in the case of the PHEME dataset, handcrafted features contributes most to both decision 
types. This implies that handcrafted features play an important role on different types of dataset generally. 
From Fig.~\ref{fig:influ_new}b, we can see that among the handcrafted features, 
ConcretenessScore (a psycho-linguistic feature), VBG (Verb,
gerund or present participle) and PDT (Predeterminer) are the most 
influential features in tweet rumor detection. Note that even though
handcrafted features contribute most in both datasets, the specific 
features differ. Hence it would indicate that in order to have the 
most general model, it is best to retain all these handcrafted features.

\begin{table*}[t]
\centering
\begin{tabular}{p{0.8cm}p{3.7cm}p{7.7cm}p{2.7cm}}
\hline
\hline
\textbf{Label }& \textbf{Source Statement}& \textbf{Top 3 Tweet Statements}  & \textbf{Attention Values } \\
\hline
\hline
\multirow{6}{0.8cm}{1}&\multirow{6}{3.7cm}{Witness: Police allegedly stopped Mike Brown after yelling at him to walk on sidewalk. \#Ferguson http://t.co/XG00R6w0k6} & \multirow{3}{8cm}{@Agent Kindi @SecretService  The \#SecretService Protects \#Obama \#PresidentObama He Get's Threats All The Time.@MichaelSkolnik}  &  0.09  \\
~&~&~&~\\
~&~&~&~\\
\cline{3-4}
~&~ & @Supreme Power @MichaelSkolnik You so edgy.&0.089 \\
\cline{3-4}
~&~ & @TimmyTurnUp @MichaelSkolnik @Supreme Power U just want to say ``white is guilty, because they white"? In Moscow black guys sold drugs...&0.076 \\
\hline
\hline  
\end{tabular}
\caption{Top three tweets (based on attention values) for the tweet record 94 in Ferguson event (PHEME dataset). A label value of 1 indicates a true rumor.
The tweet was classified correctly by the proposed model.}
\label{tab:expla}
\end{table*}

\subsubsection{Case by Case Exaplanations}
\begin{figure}[hbt!] 
\centering
\includegraphics[width=0.48\textwidth]{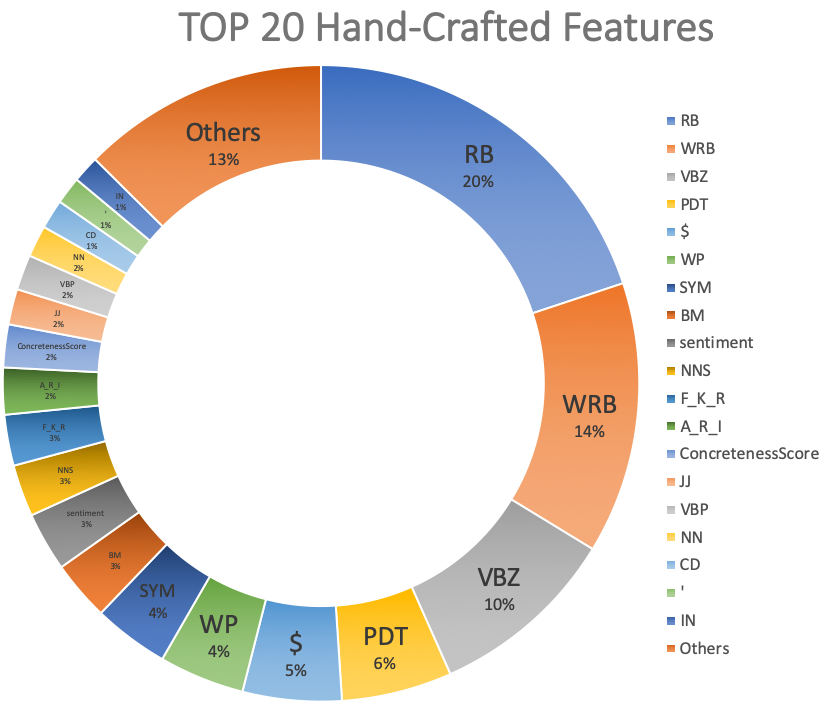}
\caption{The top 20 handcrafted features of correct detection for tweet record number 94 in Furguson sub-dataset. In this case, Unified-AttNet model identified the status of the tweet correctly.}
\label{fig:HCF}
\end{figure}
\begin{figure}[hbt!] 
\centering
\includegraphics[width=0.48\textwidth]{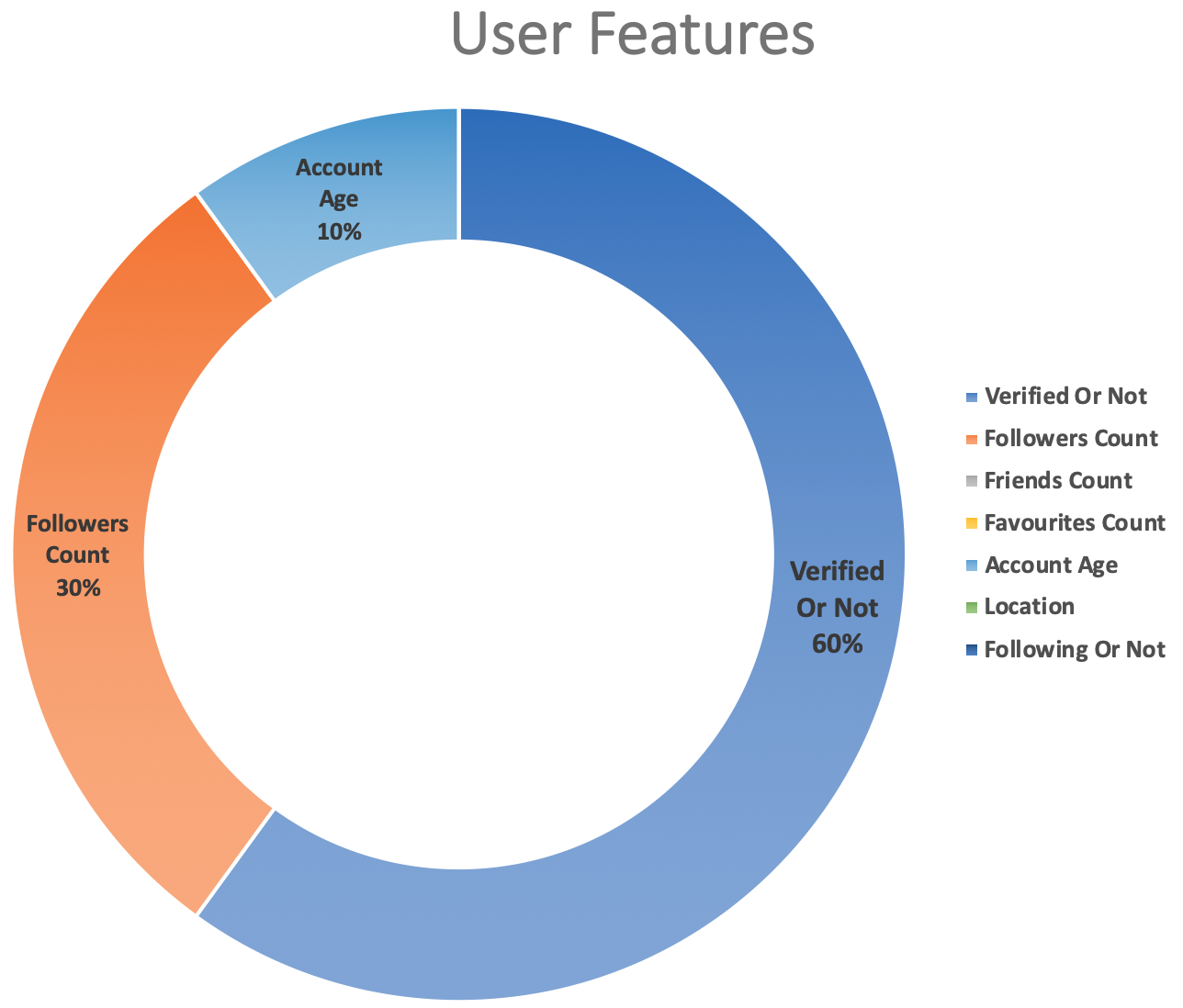}
\caption{The 7 user features of correct detection for tweet record number 94 in Furguson sub-dataset. In this case, Unified-AttNet model identified the status of the tweet correctly.}
\label{fig:UF}
\end{figure}

\begin{table}[hbt!]

\centering
\setlength{\abovecaptionskip}{0pt}%
\setlength{\belowcaptionskip}{10pt}%
\begin{tabular}{p{3.4cm}p{0.9cm}p{0.9cm}p{0.9cm}} 
\hline
\hline
\textbf{Record Number}  &857&870 & 94    \\  
\hline
\textbf{Label} & 0 &1 & 1 \\
\textbf{Latent Features}& 0.26&0.41  & 0.308 \\
\textbf{Handcrafted Features}& 0.363 &0.38 & 0.447 \\
\textbf{User Features}& 0.377&0.21  & 0.245 \\
\hline
\hline
\end{tabular}
\caption{Attention values for content latent features (USE embedding), handcrafted content features and user features for three records numbered 857, 870 and 94. Label 0 corresponds to a false rumor and Label 1 corresponds to a true rumor. In all cases, the model identified the status of the tweet correctly.}
\label{tab:dense}
\end{table}
While the previous subsection demonstrated how explanations can help in the average case for a model developer, 
in what follows, we show the interpretation of an example individual data record which may be used to explain the 
decision for one specific case to the end user.

\noindent \textbf{Explaining\ the\ role\ of\ latent\ features:}

Table~\ref{tab:expla} shows a sample tweet (correctly classified by the proposed Unified-AttNet architecture) along with its corresponding label 1 (1 denotes rumor). 
We sort the intra-feature attention values of the latent feature, pick the top 3
tweets in the conversation and display those in Table~\ref{tab:expla}. 
In this case, the top three tweets in the conversation do not include the source
tweet but rather some of the responses. In this particular example, the top three responses accounted
for $25.5\%$ of the final rumor detection result.

\noindent \textbf{Explaining\ the\ role\ of\ handcrafted\ features:}

Using intra-feature attention values we can 
interpret the relative importance of these handcrafted features towards the final decision
for the example shown in Fig~\ref{fig:HCF} where the true rumor was correctly identified by the proposed architecture.
From this figure we see that Adverb(RB), Wh-adverb(WRB) and 3rd person singular verbs in the present tense (VBZ) had the highest
values of attentions accounting for $44\%$ of the weight for handcrafted features. 

\noindent \textbf{Explaining\ the\ role\ of\ user\ features:}

An example interpretation for user features using intra-user feature
attention values is shown in Fig.~\ref{fig:UF}.
In this example, the `verified or not' and `followers count' and `account age' together account for all of the user features with $60.0\%$ weight for `verified or not', $30.0\%$ for `followers count' and $10.0\%$ for `account age'.

\noindent \textbf{Explaining\ the\ relative\ importance\ of\ feature\ classes:}

Finally we show how to interpret the relative importance between
the feature classes themselves (user features, handcrafted features and latent features) using an example. 
In Figure~\ref{fig:unified}, the last attention layer shown in the Unified-AttNet architecture
captures the relative importance among user features, handcrafted features and latent features 
in a decision. Table~\ref{tab:dense} shows three examples identified by their unique record numbers,
their corresponding grand truth labels followed by corresponding attention values for each of
of the features. For record number $94$, (identified correctly as the true rumor) the model gave 
$30.8\%$ of the weight to latent features, $44.7\%$ weight to 
handcrafted content features and $24.5\%$ weight to the user features. 
This level of explanations can give the user a better idea of what features of the tweet or news item triggered the model to arrive at its conclusion.

\section{Conclusion}
\label{sec:conlusion}

Rumors on the Internet have emerged as a modern day threat to public safety, economy and
democracy. We proposed an explainable, modular architecture for rumor
detection that can be expanded
to accommodate several feature classes, even those yet to be discovered. 
We demonstrated this architecture using
three classes of features: user features, handcrafted features derived from
content of the item and latent features obtained from
language embeddings. Using attention layers at two levels: one at the
intra-feature level for each type of feature and one at the inter-feature-class
level we achieve a granularity of explanations. The 
intra-feature level attention weights capture the relative importance that the model 
places on the individual features in the category, whereas the inter-feature 
attention weights gives us an idea of the relative importance that the 
model placed among the three classes of features. 
We also provide average case analysis of the importance of these features
which can help a model developer trim the model according to needs and
showed how to interpret the decisions on individual decisions for the end user.
Our proposed architectures perform the best among eleven benchmark models 
while providing meaningful interpretations of the decisions.

\bibliographystyle{unsrt}  
\bibliography{rumor}
\end{document}